\documentclass[usenatbib]{mnras}
\usepackage[utf8]{inputenc}
\usepackage{txfonts}
\usepackage{graphicx}
\usepackage{natbib}
\usepackage{float}
\usepackage{xcolor}
\usepackage{ulem}
\definecolor{forestgreen(web)}{rgb}{0.13, 0.55, 0.13}
\bibpunct{(}{)}{;}{a}{}{,} 
\usepackage{color}

\title
[
Granulation-generated waves and outflows
]{
Numerical experiments on granulation-generated
two-fluid 
waves and
flows
in a solar magnetic carpet 
}

\author[R.~Niedziela et al.]{
    R.~Niedziela,$^{1}$
    K.~Murawski,$^{1}$
    A.K.~Srivastava$^{2}$
\\
    $^{1}$Institute of Physics, University of M.\ Curie-Sk{\l}odowska, 
              Pl.\ M.\ Curie-Sk{\l}odowskiej 1, 20-031 Lublin, Poland\label{inst1} \\
    $^{2}$Department of Physic, Indian Institute of Technology (BHU), Varanasi-221005, India\label{inst2}
}

\begin{document}

\maketitle
    
\begin{abstract}
We consider the 
effects of 
granulation with a complex 
geometry of a magnetic carpet 
on the genesis of 
waves and plasma flows 
in a quiet-region of the solar atmosphere. 
Our aim is to perform numerical experiments 
on 
the self-generated and self-evolving solar granulation 
in a magnetic carpet representing the parts of 
the large-scale magnetized solar atmosphere, where 
waves 
and flows are basic inherent physical processes occurring continuously. 
We perform numerical 
experiments 
with the use of the JOANNA code 
which solves non-ideal and non-adiabatic two-fluid equations 
for ions+electrons and neutrals treated as two separate fluids. 
In these 
experiments, 
we assume that the plasma is hydrogen, 
and initially described by magnetohydrostatic equilibrium 
which is accompanied with a magnetic carpet. 
Parametric studies with different values of magnetic field 
show that its higher values result in larger 
magnitudes of 
ion-neutral velocity drift, 
thus ensuring larger heating and plasma flows. 
The present model addresses that in the highly dynamic solar chromosphere, waves, 
heating and 
plasma flows may collectively couple different layers of the solar atmosphere, 
and this entire process crucially depends on the local plasma and magnetic field properties.
We suggest that waves and flows are the natural response of the granulation process in the quiet-Sun.  
\end{abstract}
\begin{keywords}
Sun: atmosphere --
Sun: granulation --
methods: numerical
\end{keywords}
\section{Introduction}
It is well established 
that the plasma and radiative properties of the solar atmosphere 
change with altitude leading to an increase in its temperature 
and thus in the ionization level of the species present in its 
higher layers \citep{2008ApJS..175..229A}. 
However, the bottom layer of the Sun's atmosphere, 
called the photosphere, is characterized by a temperature range 
from about $5600$ K at its bottom to about $4300$ K at its top. 
Such low temperature leads to a weakly ionized plasma 
\citep{2014PhPl...21i2901K}.
Investigation of the small-scale magnetic activity of the three-dimensional (3D) quiet solar atmosphere models reveals mean field strength $\langle B \rangle \approx 70$~G 
in the middle of the photosphere 
\citep{2018ApJ...863..164D}. 
The photosphere is capped by the chromosphere, where temperature rises 
to almost $7\times 10^3$ K. The ionization level, therefore, 
subsequently increases and the plasma becomes partially ionized. 
The temperature in the outermost layer, known as the solar corona, 
reaches to $1 - 3$ MK, and the plasma becomes fully ionized there. 
The high-degree of rise in the temperature in upper layers 
of the solar atmosphere remains an unsolved problem, and consists of several components related to wave heating and magnetic field interactions \citep[e.g.,][]{2021JGRA..12629097S,2024arXiv240102617L}.
Such complex plasma requires a special treatment. 
One of the models, which can be used to describe weakly and partially ionized plasma, is based on the two-fluid equations \citep[e.g.][]{2011A&A...529A..82Z}. 
This model naturally 
takes into account 
ion-neutral collisions \citep{2018SSRv..214...58B} 
which result in wave damping and consequently in wave energy thermalization \citep{2007A&A...461..731F, 2004A&A...427.1055E}. 
The last studies of the two-fluid waves performed by 
\cite{2022A&A...668A..32N}, \cite{2022Ap&SS.367..111M} and 
\cite{2023A&A...669A..47P} show their contribution to 
chromosphere heating and generation of plasma outflows in the low corona.

It is likely that 
convective movements of plasma under the solar surface lead to 
the formation of solar granulation, and they can be a source of 
many dynamical events 
and excitation of waves \citep[e.g.][]{2017ApJ...835..148V}.

In this context, convective cells and 
magnetic field, together form 
magneto-convection \citep{2004A&G....45d..14P}. 
The generation and propagation of magnetoacoustic waves due to granulation was widely studied \citep[e.g.][]{2006ApJ...647L..73H, 2011ApJ...743..142H}. Additionally, \cite{2017Sci...356.1269M} investigated excitation of solar spicules and Alfv\'en waves in the 2.5-dimensional (2.5D) model with solar granulation. 
Recent studies of the two-fluid waves generated by spontaneously generated and self-evolving convection show that a wide spectrum of wave periods is generated by the granulation 
\citep{2020A&A...635A..28W}. In the later studies, \cite{2021RSPTA.37900170F} performed numerical simulations of acoustic-gravity waves generated by the solar granulation and confirmed that only short-period acoustic and long-period gravity waves are able to reach the corona.

Along similar lines, 
\cite{2022Ap&SS.367..111M} studied chromosphere heating and generation of plasma outflows associated with two-fluid 
solar granulation. However, initially (at $t = 0$ s) straight vertical magnetic field was considered. 
The major aim of this paper is to extend the model of 
\cite{2022Ap&SS.367..111M} by supplementing the vertical magnetic field 
by a more realistic solar magnetic carpet which naturally occupies 
the solar atmosphere 
in the form of magnetic arcades \citep{2002MNRAS.335..389P}. 
Such magnetic carpet is expected to significantly affect 
the dynamics of the localized solar atmosphere and, as a result, 
influences chromosphere heating and plasma flows inherent therein. 

In the present paper, we illustrate 
the comperehensive physical scenario of 
the self-consistent evolution of waves and flows excited by 
the granulation, operating in 
the 
two 
regimes of 
magnetic carpet. We also emphasize on the dependence of 
these physical processes on the intensity of the magnetic field. 

This paper is organized as follows. In the following section, 
the numerical model is described. Section 3 presents the results of 
the numerical 
experiments. 
The last section contains the summary and conclusions.
\section{Numerical model}
%
In this paper, we consider the solar atmosphere which consists of 
partially ionized hydrogen plasma which 
dynamics can be described by the set of non-ideal and non-adiabatic two-fluid equations with operating ionization and recombination for ions (protons) + electrons and neutrals (hydrogen atoms). 
These equations and the adopted numerical methods were described 
in detail by \cite{2022Ap&SS.367..111M}. 
Here, we 
use the extra heating term which is equal to $99$ \% 
of the thin radiation, and 
limit ourselves to a presentation of all 
the necessary information only that are utilized in the numerical setup. 

\begin{figure}
\centering
\includegraphics[width=8cm]{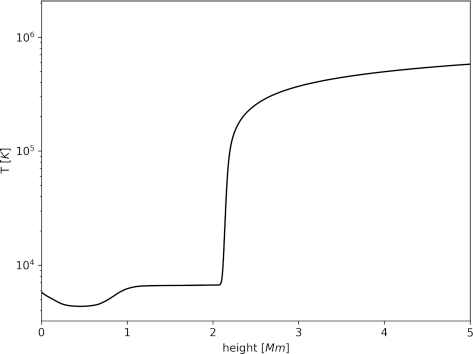}
\includegraphics[width=8cm]{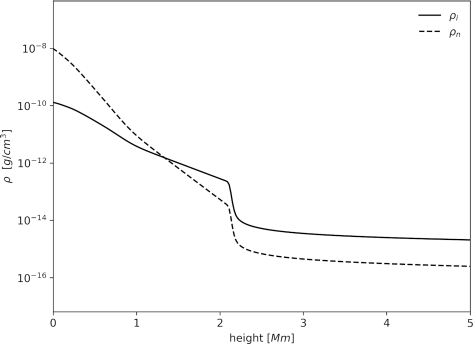}
\caption{Variation with height, $y$, of the initial temperature, 
$T_{\rm 0}$, (top) and 
ion (solid line) 
and 
neutral (dashed line) 
mass densities 
(bottom).
}
\label{fig:equilibrium}
\end{figure}

Initially (at $t = 0$ s), we set the magnetohydrostatic equilibrium 
with hydrostatic ion and neutral gas pressures and 
mass density profiles \citep{2022Ap&SS.367..111M} 
which result from the semi-empirical temperature profile, 
$T_{\rm 0}(y)$, 
of \citep{2008ApJS..175..229A}. See Fig. \ref{fig:equilibrium} 
for the temperature (top) and mass density (bottom) profiles. 
Note that ion mass density, $\varrho_{\rm i}$, is about 
$100$ times smaller than neutral mass density, $\varrho_{\rm n}$, 
at $y=0$ Mm. At $y\approx 1.3$ Mm 
$\varrho_{\rm i}=\varrho_{\rm n}$ and higher up in the chromosphere and in the solar corona, 
$\varrho_{\rm i}$ becomes larger than $\varrho_{\rm n}$. 
The hydrostatic profiles are 
overlaid by the arcade magnetic field that is given as 
\begin{equation}
\mathbf{B} = B_{\rm a} \left[\cos \left(\frac{x+L_{\rm B}}{\Lambda_B}\right), 
-\sin \left(\frac{x+L_{\rm B}}{\Lambda_B}\right), 0\right] e^{-y/\Lambda_B} + \left[0, B_{\rm v}, B_{\rm t}\right].
    \label{eq:magnetic}
\end{equation}
Here, $B_{\rm a}$, $B_{\rm v}$ and $B_{\rm t}$ correspond respectively to a magnetic carpet, modelled by the set of arcades, 
vertical and transversal components of magnetic field, $L_{\rm B} = 0.64$ Mm is the half-size of a single arcade, and 
$\Lambda_B = 2 L_{\rm B} / \pi$ denotes a height over which $B$ falls off $e-$times. 

Figure \ref{fig:eq} (top) illustrates 
spatial profiles of 
$T_{\rm i}(x,y,t=0\, {\rm s})$ and 
magnetic field lines which correspond to Eq.~(\ref{eq:magnetic}) 
with $B_{\rm a}=0.2$~G, and are set initially, at $t=0$~s. 
It should be noted that the magnetic carpet is located 
below the transition region that is 
initially set 
at $y \approx 2.1$ Mm. 
As a result of the implementation of the vertical magnetic field, $B_{\rm v} = -5$~G, there are seventeen magnetic null points 
(Fig.~\ref{fig:eq}, top). The transverse component of magnetic field is chosen as $B_{\rm t} = 1$~G. 
This choice of $B_{\rm v}$ and $B_{\rm t}$ is appropriate 
for the upper chromosphere and the corona. 

To solve the two-fluid equations numerically, we use the JOANNA code \citep{2020A&A...635A..28W}. 
Along the $y$-direction we covered the numerical domain with $512$ cells in the fine grid zone which occupies the region 
$(-3.0 \leq y \leq 17.48)$ Mm. Higher up, the grid is stretched and consists of $16$ cells up to $y=25$ Mm. 
The size of the numerical box along the $x-$direction is ($-20.48 \leq x \leq 20.48$) Mm and it is covered by $1024$ cells, 
leading to the finest grid resolution of 
$\Delta x = \Delta y = 40$ km. 
The plasma quantities are maintained at their equilibrium values at the top and bottom boundaries of the numerical box, while at the left- and right-sides periodic boundary conditions are implemented.
%
\section{Numerical results}
In this section, we present the dynamics of the model solar atmosphere 
in two regimes of the magnetic carpet, namely 
$B{\rm a}=0.2$~G and $B{\rm a}=0.075$~G.
\subsection{The case of \texorpdfstring{$B_{\rm a}=0.2$}~G} 
%
%
\begin{figure}
\centering
\includegraphics[width=8cm]{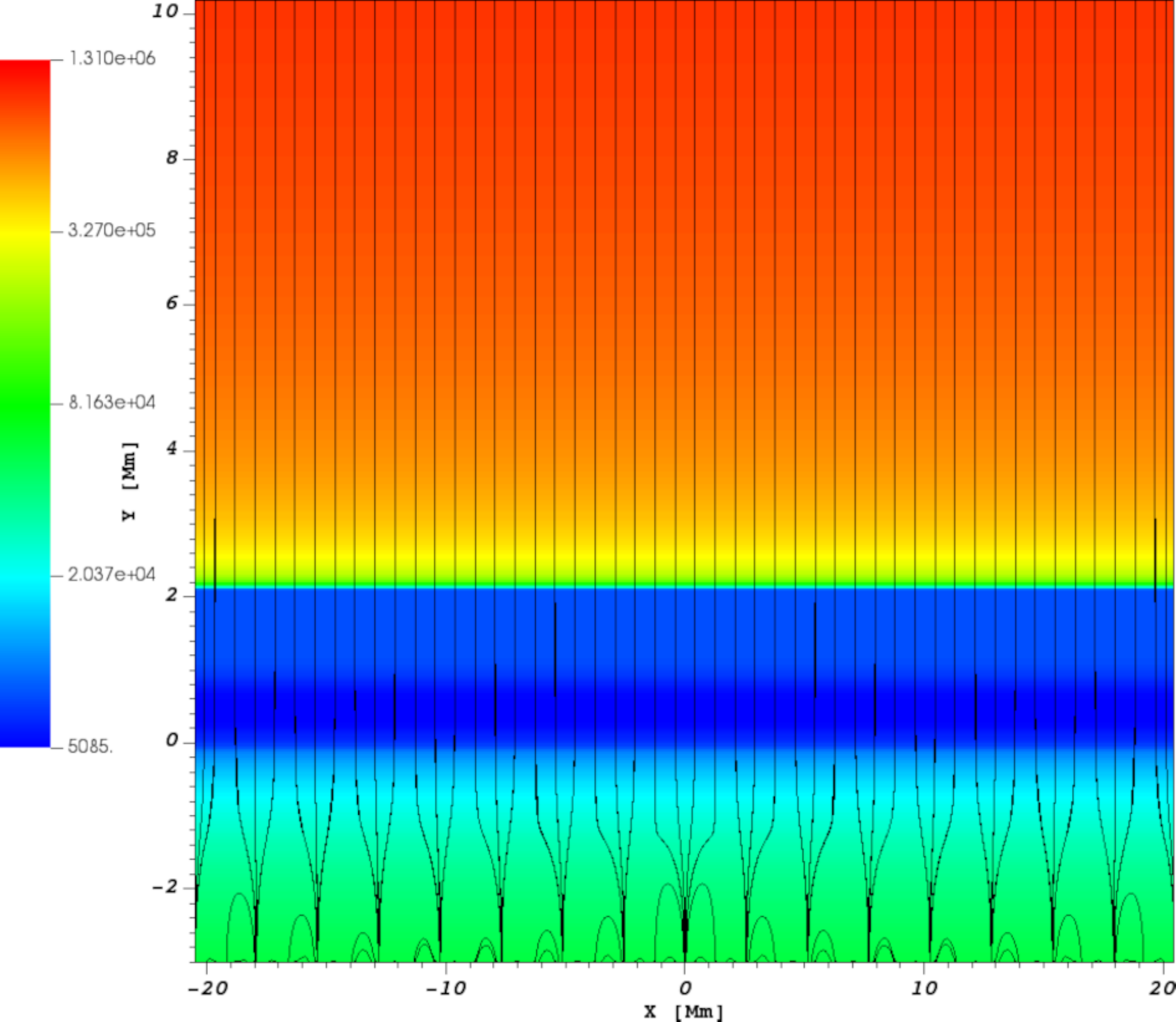}
\includegraphics[width=8cm]{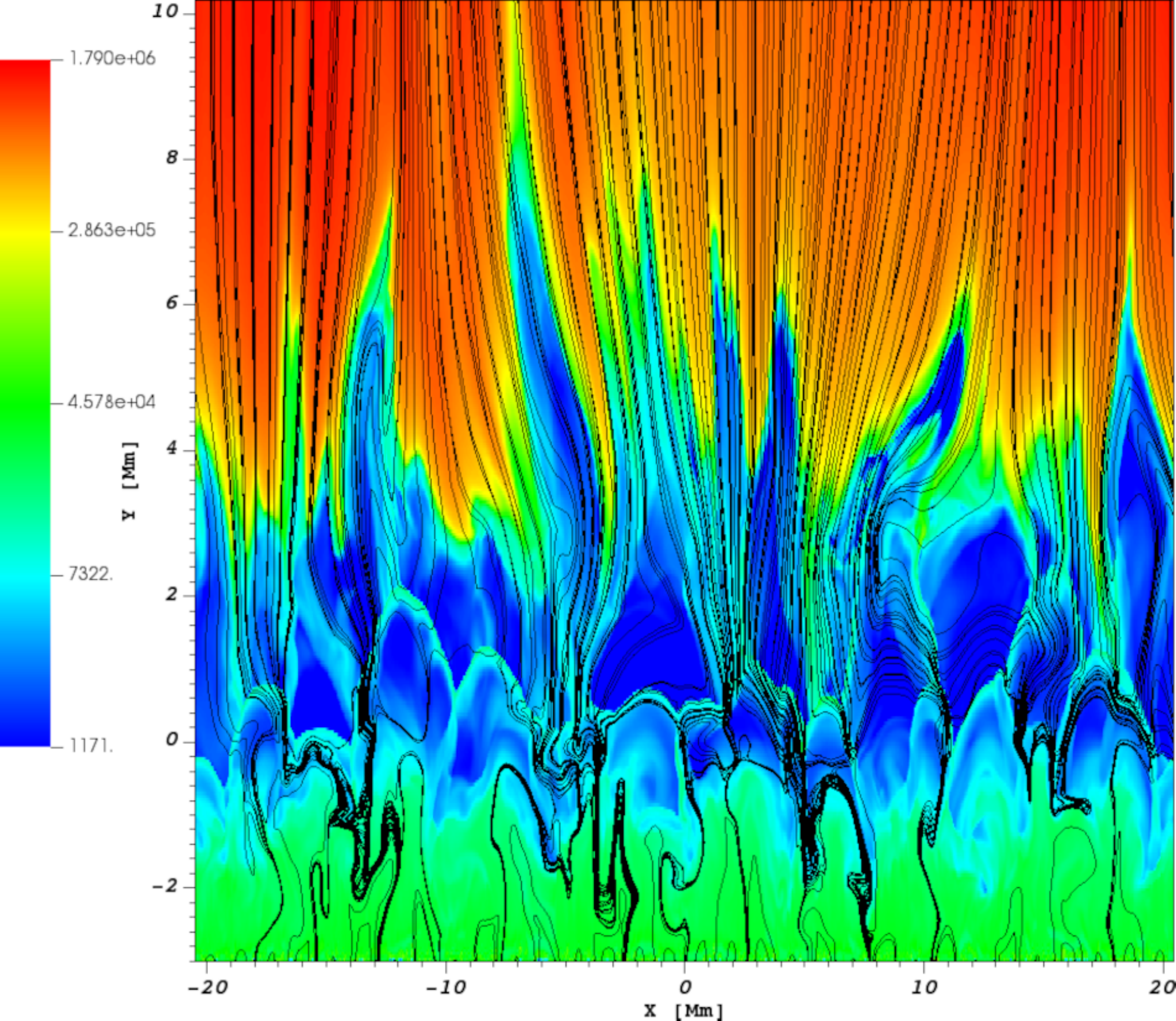}
\caption{Spatial profiles of log($T_{\rm i}$) at $t = 0$ s (top) and $t = 5000$ s (bottom), overlaid by magnetic field lines 
which correspond to a magnetic carpet with $B_{\rm a} = 0.2$~G.}
\label{fig:eq}
\end{figure}
%
%
%
\begin{figure*}
\centering
\includegraphics[width=8cm]{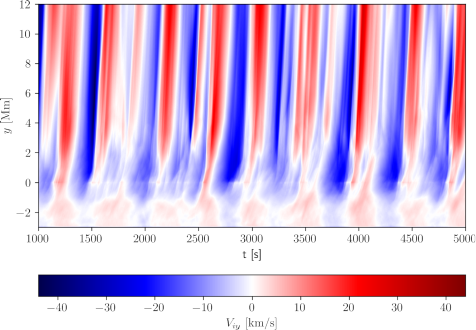}
\includegraphics[width=8cm]{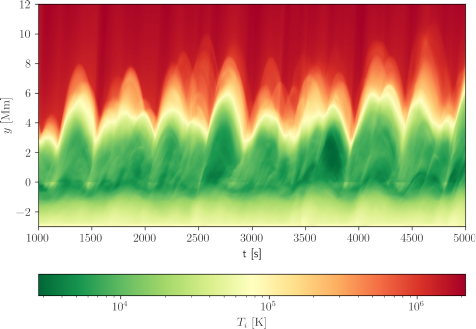}
\includegraphics[width=8cm]{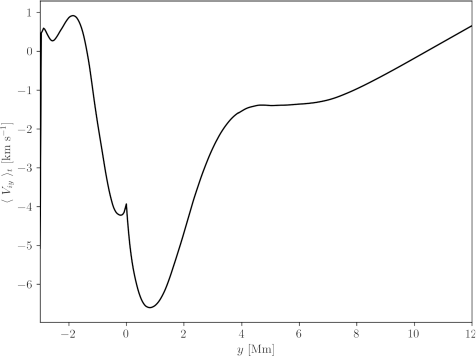}
\includegraphics[width=8cm]{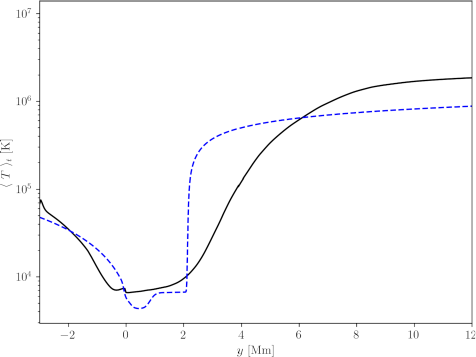}
\caption{Time-distance (top) and averaged over time (bottom) plots 
for horizontally averaged $\langle V_{\rm iy}\rangle_{\rm x}$ (left) 
and $\langle T_{\rm i}\rangle_{\rm x}$ (right) (solid lines) for $B_{\rm a}= 0.2$~G, $B_{\rm v}= -5$~G and $B_{\rm t}= 1$~G.
Dashed line corresponds to \protect\cite{2008ApJS..175..229A} temperature profile (bottom right).
}
\label{fig:Viy_Ti}
\end{figure*}
We display results here for $B_{\rm a}= 0.2$~G. 
See Eq. (\ref{eq:magnetic}). 
The waves are excited by the self-generated and self-evolving solar granulation 
which reshuffles magnetic field lines 
particularly strongly at the bottom of 
the photosphere and below it. 
The granulation mechanism takes place at 
the bottom of the photosphere, 
where convective instabilities lead to the formation of 
the turbulent flows there.
These flows are associated with the perturbation of 
the initial state (Fig.~\ref{fig:eq}, top) and thus  
the ejection of ions and neutrals to the higher layers of 
the atmosphere. 
The jets shown on the spatial profile of $\log(T_{\rm i})$ (Fig.~\ref{fig:eq}, bottom) 
are generated by the solar granulation 
which leads to reconnection of magnetic field lines 
and thermal energy release in the photosphere and 
the chromosphere. 
This release in the chromosphere results in 
the largest jet which arrives to a height of about $y = 10$ Mm 
and it is located at $x = -6$ Mm. 

In a progress of time the magnetic carpet evolves into complex magnetic arcades in the upper atmospheric layers.
The jets obtained in the numerical 
experiments for $B_{\rm a}= 0.2$~G 
reach higher altitudes 
than in the initially straight magnetic field system 
which was considered by \cite{2022Ap&SS.367..111M}. 
At the foot-points of this carpet, that is below 
$y = 0$ Mm, magnetic field lines form the small magnetic flux-tubes of $B \approx 1332$~G and 
with strong downflows \citep{2022Ap&SS.367..111M}. 

In the presentation of the numerical results, 
we use below averaged plasma quantities: 
%
\begin{eqnarray}
    \langle f \rangle_{\rm x}  &=& 
    \frac{1}{x_{\rm 2} - x_{\rm 1}} 
    \int_{x_{\rm 1}}^{x_{\rm 2}} f \; dx \, ,
    \label{eq:f_x} \\
    \langle f \rangle_{\rm xt}  &=& 
    \frac{1}{t_{\rm 2} - t_{\rm 1}} 
    \int_{t_{\rm 1}}^{t_{\rm 2}} \langle f \rangle_{\rm x} \; dt \, .
    \label{eq:f_xt} 
\end{eqnarray}
Here, 
$t_{\rm 1} = 1000 \; \rm{s}$, 
$t_{\rm 2} = 5000 \; \rm{s}$, $x_{\rm 2} = -x_{\rm 1} = 20.48$ Mm, and 
$f$ is a plasma quantity such as 
a horizontally averaged vertical component of 
ion velocity, $V_{\rm iy}$, 
and relative ion temperature perturbations, 
\begin{equation}
\Delta T_{\rm i} = \frac{T_{\rm i} - T_{\rm 0}}{T_{\rm 0}}\, .
\end{equation}
%


Plasma motions are in the form of upflows and downflows 
which are accompanied by two-fluid ion magnetoacoustic-gravity, 
neutral acoustic-gravity, and Alfv\'en waves.
Maximum value of the averaged ion velocity, 
$\left<V_{\rm iy}\right>_{\rm x}$, 
reaches about $44 \; \rm{km\:s^{-1}}$ 
(Fig.~\ref{fig:Viy_Ti}, top-left). 
Note that $\left<V_{\rm iy}\right>_{\rm xt}$ attains its minimum 
of about $-6.5$ km s$^{-1}$ at $x=1$ Mm. 
Higher up $\left<V_{\rm iy}\right>_{\rm xt}$ grows with $y$ 
and at $y\approx 10$ Mm it reaches its positive values, 
resulting in a net plasma outflows in the corona. 
See Fig.~\ref{fig:Viy_Ti} (bottom-left). 

The solar granulation results in ejection of ions and neutrals
from the chromosphere, 
which while arriving to the transition region 
result in its oscillations. See Fig.~\ref{fig:Viy_Ti} (top-right), 
illustrating time-distance plots for 
$\langle T_{\rm i} \rangle_{\rm x}$. 
Note that the transition region bounces up and down with 
oscillations which progressively 
calm down 
in time. 

\begin{figure}
\centering
\includegraphics[width=8cm]{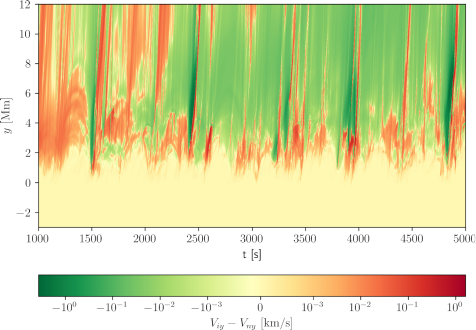}
\caption{Time-distance plot for the horizontally averaged vertical component of 
ion-neutral velocity drift, $\left<V_{\rm iy}-V_{\rm ny}\right>_{\rm x}$, 
in the case of $B_{\rm a} = 0.2$~G.
}
\label{fig:Viy_Vny}
\end{figure}
%
As a consequence of ion-neutral collisions, 
the excited waves are dissipated. 
This effect is most effective at the places, 
where the difference between the velocities of ions and neutrals is being 
the largest \citep{2020ApJ...900..101M}. 
See Fig.~\ref{fig:Viy_Vny} which illustrates the vertical component of 
the ion-neutral velocity drift, $\left<V_{\rm iy} - V_{ny}\right>_{\rm x}$. 
This drift attains its largest values in the 
transition region and low corona, 
which is in a good agreement with \cite{2022Ap&SS.367..111M}.
It results from the weak coupling of ions and neutrals in the higher layer of the solar atmosphere and strong coupling in the lower regions. Thus they propagate with essentially the same speed in the photosphere 
and the chromosphere.
Note that in the 
low corona 
during the initial phase 
$\left<V_{\rm iy} - V_{ny}\right>_{\rm x} > 0$. 
Hence at $t < 1500$ s, as a result of the Lorentz force 
acting on them, 
ions attain higher velocities 
than neutrals which are not directly affected by this force.  

The horizontally and time-averaged relative ion temperature, 
$\langle \Delta T_{\rm i} \rangle_{\rm xt}$, 
illustrates lower values compared to the semi-empirical data of \cite{2008ApJS..175..229A} in the upper part of the convection zone and lower corona. Note that the horizontally and temporally averaged velocity reaches negative values, $\left<V_{\rm iy}\right>_{\rm xt} < 0$, below the altitude $y = 6$ Mm. However, above this level, $\left<V_{\rm iy}\right>_{\rm xt}$ attains positive values, which reveals the net plasma outflows. 
A comparison with \cite{2022Ap&SS.367..111M} findings shows that the maximum $\left<V_{\rm iy}\right>_{\rm x}$ values in both models are similar. 
But our results show a higher contribution of plasma outflows compared to plasma downflows with the opposite trend reported by \cite{2022Ap&SS.367..111M}. This is a significant new aspect as evident in the system of magnetic carpet with $B_{\rm a}= 0.2$~G. 
Besides, the simplified model of radiative losses, 
used in our model, 
may lead to the flows at the transition region and to its smoothing 
as it is evident in the averaged temperature profile in 
Fig.~2 (bottom-right panel). 

\begin{figure}
\centering
\includegraphics[width=8cm]{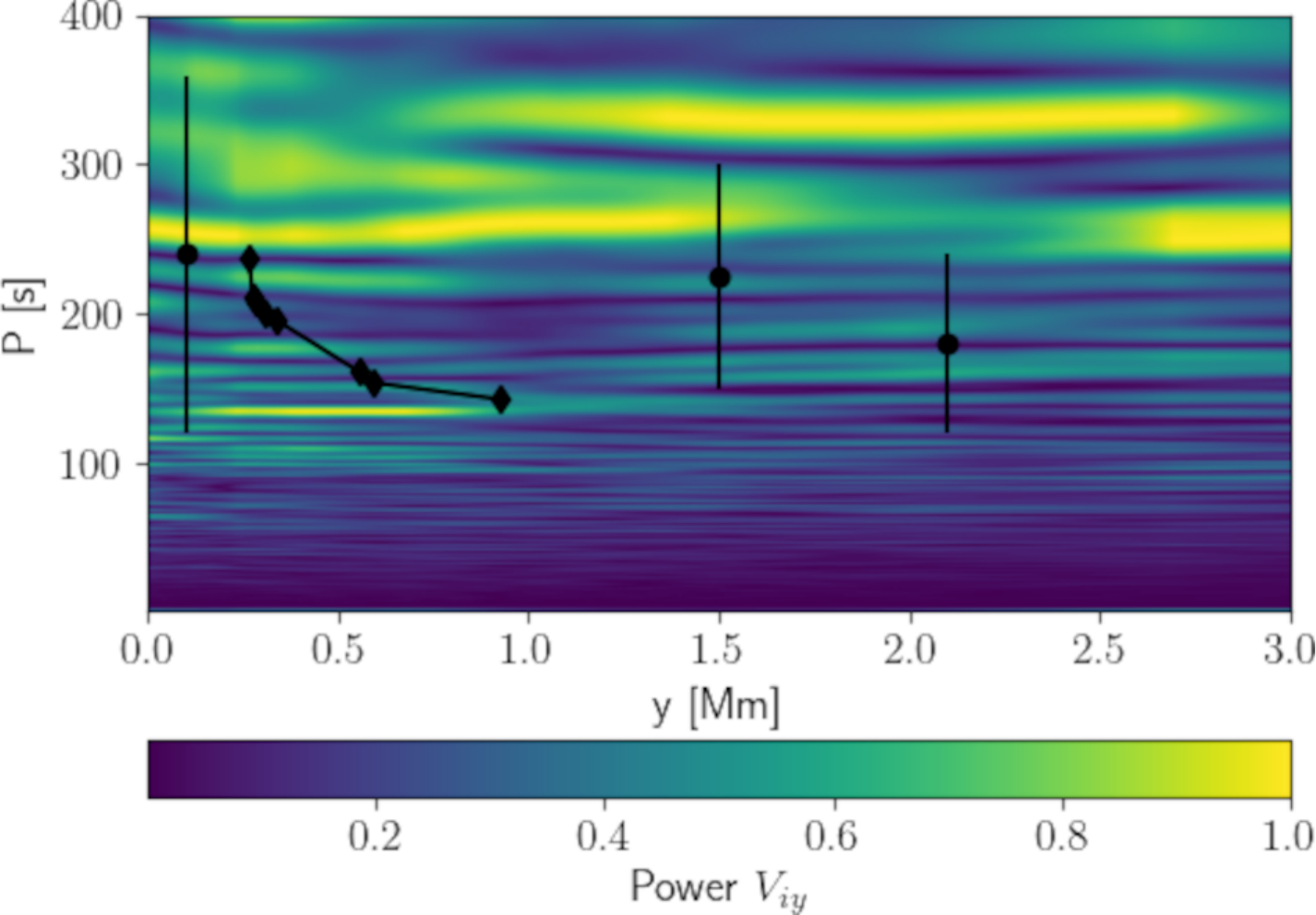}
\caption{Fourier power spectrum of wave period $P$ for 
$\langle V_{\rm iy}\rangle_{\rm x}$ vs. height 
for $B_{\rm a}=0.2$~Gs. 
The diamonds and dots correspond to the observational data obtained by 
respectively \protect\cite{2016ApJ...819L..23W} and \protect\cite{2018MNRAS.479.5512K}.
}
\label{fig:Periods}
\end{figure}
%
%
%
%
\begin{figure}
\centering
\includegraphics[width=8cm]{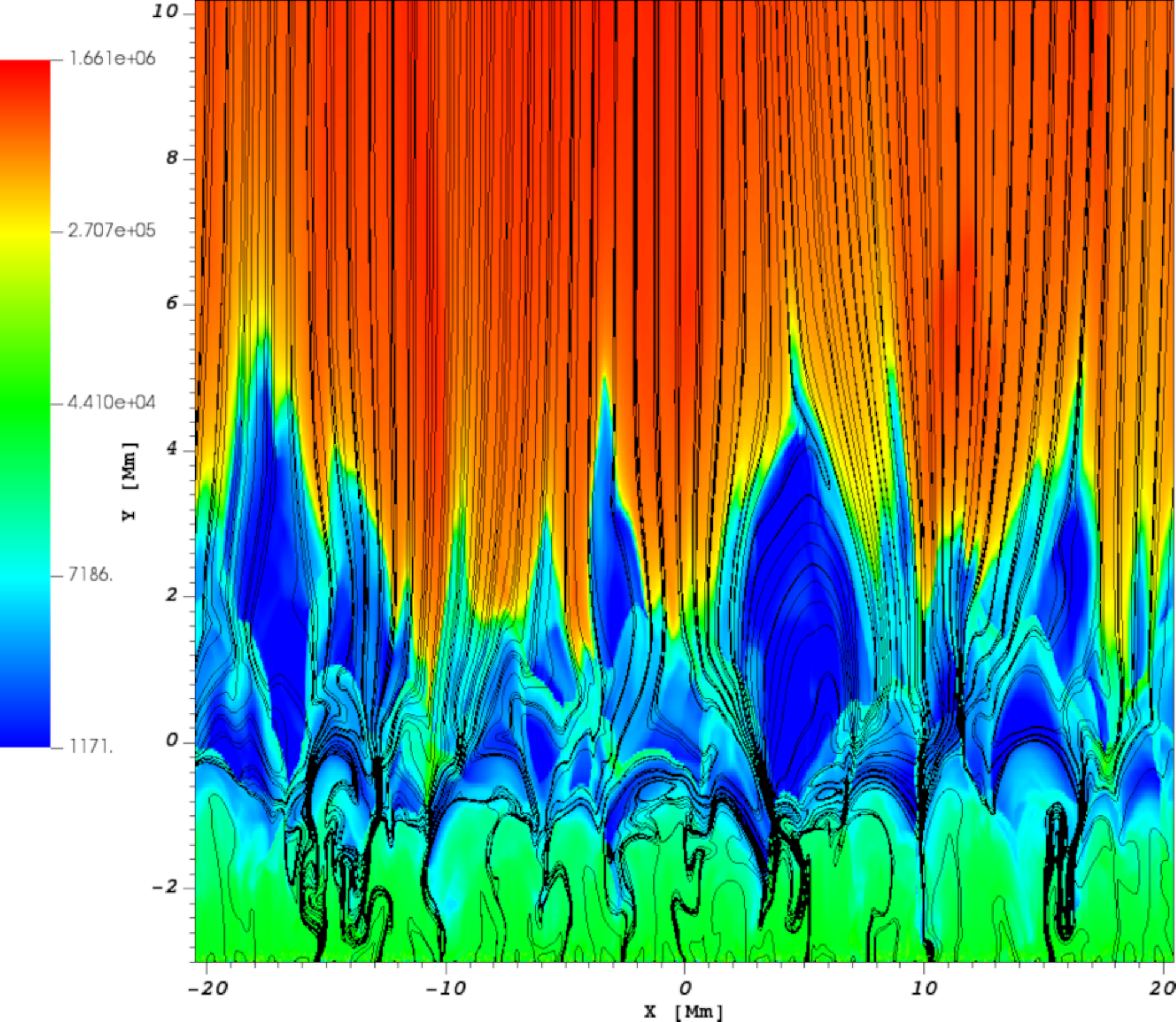}
\caption{Spatial profiles of log($T_{\rm i}$) at $t = 5000$ s, 
overlaid by magnetic field lines which correspond to a magnetic carpet for $B_{\rm a} = 0.075$~G.}
\label{fig:spatial_weaker}
\end{figure}
In stratified medium such as the solar atmosphere, cutoff periods is an important quantity that determines the wave period ranges 
below which waves are able to propagate upwards 
\citep[e.g.][]{2020Ap&SS.365..139R}. 
Figure \ref{fig:Periods} illustrates wave periods obtained from 
the Fourier spectra for 
$\left<V_{\rm iy}\right>_{\rm x}$ 
(Fig.~\ref{fig:Viy_Ti}, top-left). 
We observe that at the heights within 
the range $0 \: \rm{Mm} < y < 1.5$ Mm, 
the main wave period is about $P = 250$ s. 
For $y > 1.5$ Mm, the dominant wave period corresponds to 
$P \approx 340$ s. 
It is well known that waves 
with a period $P \approx 300$ s are evanescent in the photosphere 
\cite[e.g.][]{2018MNRAS.481..262W} and therefore waves of these 
wave periods are unable to extend to higher altitudes. 
A comparison with \cite{2022Ap&SS.367..111M} reveals that in both 
cases, we observe multiple wave power concentrations for various 
periods and heights. However, our results show weaker agreement with the observational data.
 
%
%
\subsection{The case of \texorpdfstring{$B_{\rm a}=0.075$}~G}
We present results here for $B_{\rm a}=0.075$~G. 
Figure~\ref{fig:spatial_weaker} displays spatial profiles of 
$\log(T_{\rm i})$ (color maps) and magnetic field lines 
at $t=5000$~s 
for $B_{\rm a} = 0.075$~G. 
There is a clear difference in the size of the jets: those from 
the magnetic carpet with $B{\rm a}=0.075$~G 
reach an altitude of $y = 5$ Mm, while jets from $B{\rm a}=0.075$~G reach $y = 10$ Mm.
Additionally, we observe that the higher $B_{\rm a}$ value affects higher 
maximum temperature of ions.

\begin{figure*}
\centering
\includegraphics[width=8cm]{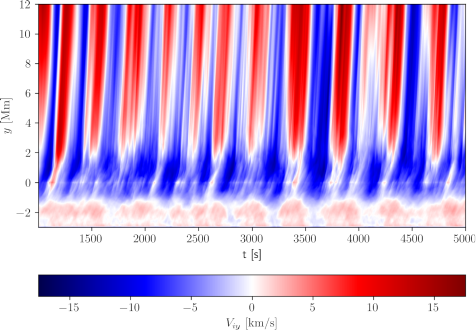}
\includegraphics[width=8cm]{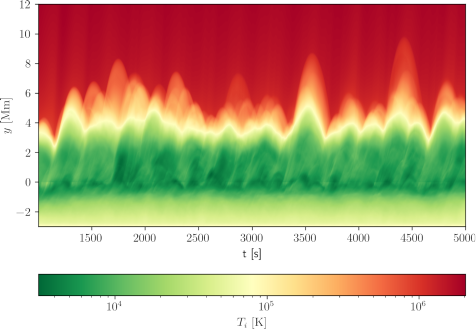}
\includegraphics[width=8cm]{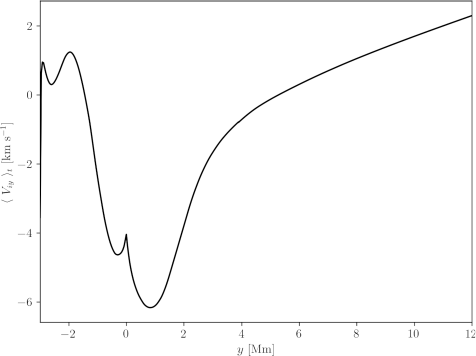}
\includegraphics[width=8cm]{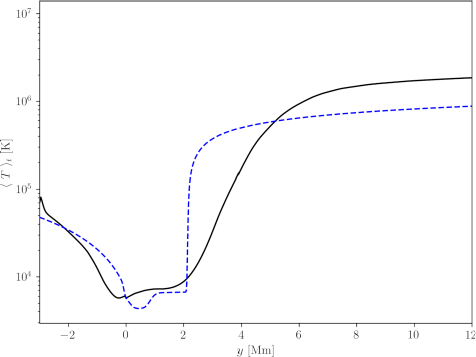}

\caption{Time-distance (top) and averaged over time (bottom) plots 
for horizontally averaged $\langle V_{\rm iy}\rangle_{\rm x}$ (left) 
and $\langle T_{\rm i}\rangle_{\rm x}$ (right) for $B_{\rm a}= 0.075$~G.
}
\label{fig:Viy_Ti_Ba}
\end{figure*}
Figure~\ref{fig:Viy_Ti_Ba} illustrates time-distance plots for 
vertical component of ion velocity, 
$\left<V_{\rm iy}\right>_{\rm x}$, (top-left) 
and ion temperature, $\left<T_{\rm i}\right>_{\rm x}$, (top-right), averaged over the horizontal direction. 
From the magnetic carpet studies, we infer that 
increasing $B_{\rm a}$ from 
$B_{\rm a}= 0.075$~G to $B_{\rm a}= 0.2$~G affects significantly max($\left<V_{iy}\right>_{\rm x})$ 
which increases from about $19 \; \rm{km\:s^{-1}}$ for 
the arcade magnetic field 
of $B{\rm a}=0.075$~G
to about $43 \; \rm{km\:s^{-1}}$ for $B_{\rm a}=0.2~G$. 
Such increment of the ion vertical flows is anticipated 
as a stronger magnetic field corresponds to 
a much larger Lorentz force.
Additionally, we observe change in the height 
that the oscillations of the transition region reach which is higher 
for 
$B_{\rm a}=0.2$~G.
For the two $B_{\rm a}$ values studied in this article, we observe plasma outflows in the corona, while the photosphere and the chromosphere are dominated by the downflows. 
However, $\left<V_{iy}\right>_{\rm xt} > 0$ are present in the corona for higher altitudes for 
the magnetic carpet with 
$B_{\rm a}=0.2~G$, viz. $y \approx 10$ Mm, than for 
the magnetic carpet with 
$B_{\rm a}=0.075~G$, 
$y \approx 5.5$ Mm (Fig.~\ref{fig:Viy_Ti_Ba}, bottom-left). 

\begin{figure}
\centering
\includegraphics[width=8cm]{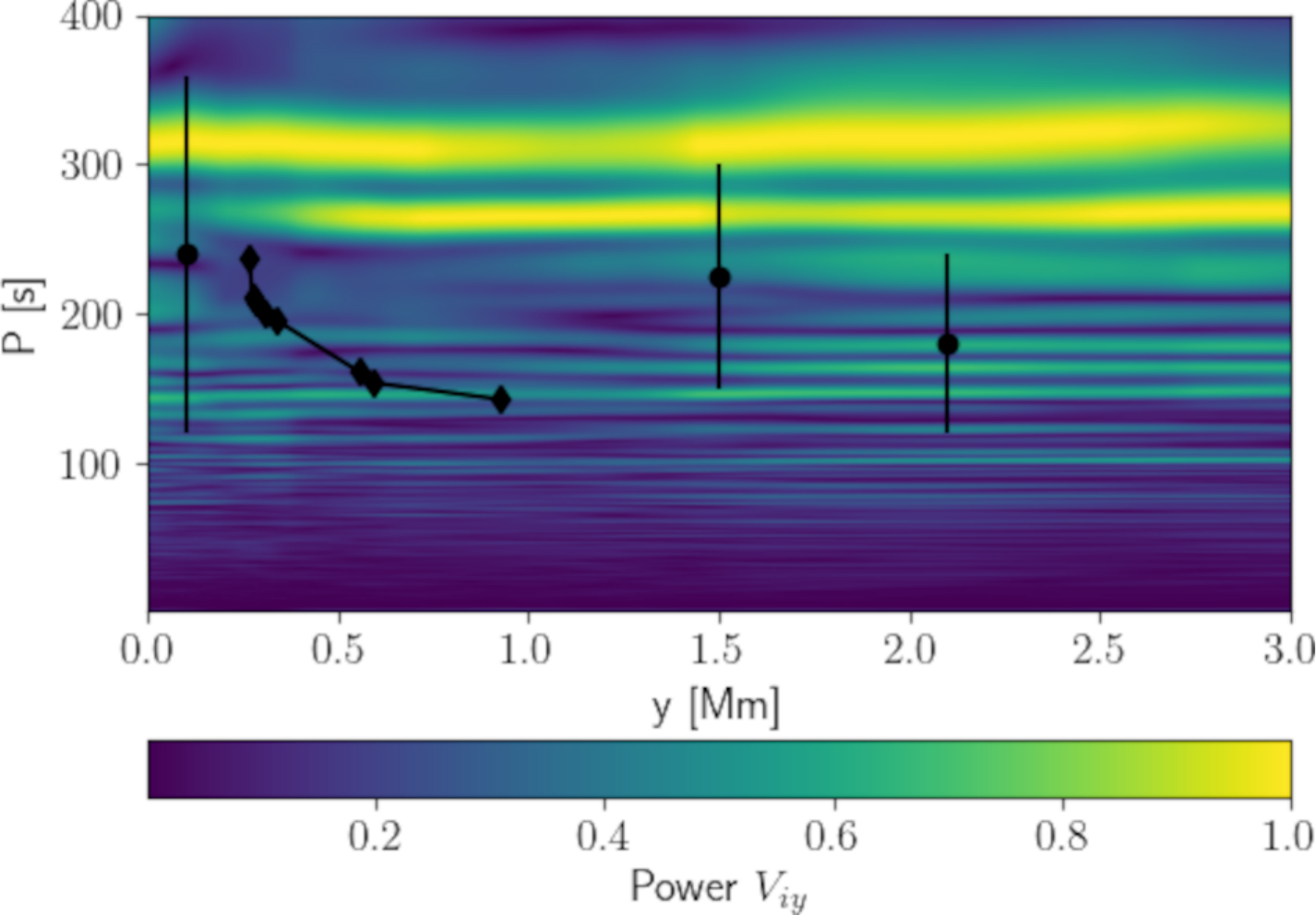}
\caption{Fourier power spectrum of wave period $P$ for $\left<V_{\rm iy}\right>_{\rm x}$ vs. height for $B_{\rm a} = 0.075$~G. 
The diamonds and dots correspond to the observational data obtained by \protect\cite{2016ApJ...819L..23W} and \protect\cite{2018MNRAS.479.5512K}. 
}
\label{fig:Periods_weak}
\end{figure}

Figure \ref{fig:Periods_weak} presents wave period, $P$, 
vs. height, obtained from the Fourier power spectra for 
$\left<V_{\rm iy}\right>_{\rm x}$ for 
$B_{\rm a}=0.075$~G. 
In this case, 
we can distinguish two dominant periods, namely  
$P \approx 320$ s (observed in 
for $B_{\rm a}=0.2$~G 
for $y>0.7$ Mm) and $P \approx 270$ s, illustrated by 
the two yellow strips. 
The former period is surprisingly present essentially 
at every height; 
such long period waves are anticipated to be evanescent 
as periods higher than the cutoff period correspond 
to non-propagating waves \citep{kuzma2024RSPTA.38230218K}. 
As a result, we infer that the plasma background 
is altered in time, increasing the cutoff period and 
allowing so large period waves to propagate 
from the photosphere through the chromosphere into 
the corona. 
The latter period is seen at $y\approx 0.5$ Mm 
and higher up, which evidences that such period waves 
propagate freely into the corona. 
Besides these two major wave periods, shorter periods 
waves with $P$ being within the range of 
about $100-200$ s are also generated by the granulation 
and they are seen throughout the whole atmosphere. 
For 
$B_{\rm a}=0.2$~G 
$P\approx 250$ s 
does not show up for $1.5\:~\rm{Mm} < y < 2.5$~Mm. 
Comparison with the observational data of 
\cite{2016ApJ...819L..23W} (diamonds) and \cite{2018MNRAS.479.5512K} (dots) 
reveals some level of agreement 
at certain altitudes. 
%
\section{Summary and conclusions}
We performed 2.5 D numerical 
experiments 
of the 
solar atmosphere 
that is modelled by two-fluid equations with 
non-adiabatic, non-ideal effects and ionization/recombination effects 
taken into account. We aimed to investigate 
chromosphere heating and generation of plasma outflows 
as well as Fourier power spectrum of 
the excited waves. 

Our results can be summarized as follows. 
The developed numerical model showed that the self-generated and self-evolving solar granulation, 
along with the considered magnetic field configuration in the form of 
a magnetic carpet, facilitates the generation of all two-fluid waves, heating and flows. 
The study considered 
magnetic carpets with 
$B_{\rm a} = 0.2$~G and $B_{\rm a} = 0.075$~G. 
Given 
the observations (e.g. via the Hanle effect) 
which indicate a field strength of about $70$~G 
in the mid-photosphere, 
these values are initially (at $t=0$ s) small. 
However, the self-generated and self-evolved granulation 
alters the magnetic field which is organized in flux-tubes 
with mximum magnitudes of magnetic field of about 
$\mathbf{B} = 1332$~G and 
$\mathbf{B} = 650$~G for 
$B_{\rm a}=0.2$~G and $B_{\rm a}=0.075$~G, 
respectively. 
From the obtained numerical results, we infer that 
energy carried to the upper atmospheric layers 
thermalize and, 
as a result of the ion-neutral collisions, leading to localized heating of 
the chromosphere. 
Note that for both $B_{\rm a}$, $\left<T_{\rm i}\right>_{\rm xt}$ reaches higher values than the semi-empirical temperature model of \cite{2008ApJS..175..229A} in the entire photosphere and chromosphere and the only distinguishable difference occurs in the corona. 
Nevertheless, the numerical findings of \cite{2022Ap&SS.367..111M} show better agreement with semi-empirical temperature data of \cite{2008ApJS..175..229A} than in our case (Fig~\ref{fig:Viy_Ti}, right-bottom). 
We speculate that 
it may be caused by the use of the different values 
of the extra heating implied in the corona,  
mainly, in our case  
the extra heating balanced $99$ \% of the thin cooling  
and $100$ \% in \cite{2022Ap&SS.367..111M}.

The 
resulting 
plasma heating is 
accompanied by plasma outflows. 
Hence, all the two cases, we consider here, can be the source of the nascent solar wind due to certain degree of plasma outflows generated in the corona. 
At higher altitudes, 
the 
magnetic carpet in the quiet-Sun effectively transport the plasma contributing to the origin of the nascent solar wind and mass cycle 
in the solar atmosphere. 
This result converges with the finding of 
\cite{2005Sci...308..519T} and \cite{2010ApJ...709L..88T} 
who suggest that the solar wind originates from coronal funnels 
above $y = 5$ Mm. 

Studies of the 
magnetic carpets, specified by initially smaller and larger 
values of magnetic fields, $B_{\rm a}$, show clear differences in 
the size of the generated jets and 
plasma oscillations. 
Additionally, we observe increase of the plasma velocity for higher $B_{\rm a}$, 
which is anticipated as 
larger vertical ion flow may result from higher Lorentz force.
The difference between the solutions corresponding to the different values of $B_{\rm a}$ 
is supposed to disappear after a long time. 
These solutions are anticipated to become statistically indistinguishable 
as time goes to infinity, and the differences present at any finite time 
represent transients of the initial condition.


The results obtained for the wave periods exhibit 
dominant periods 
and reveal 
some level of agreement with the observational data 
of \cite{2016ApJ...819L..23W} 
and \cite{2018MNRAS.479.5512K} 
at 
some altitudes. 

While the developed model 
mimics some aspects of 
the excited wave spectrum, 
the chromosphere heating and generation of plasma outflows, 
they do not show the whole scenario, 
and a more sophisticated treatment is required. 
Taking into account additional mechanisms to describe the solar atmosphere more accurately is a major challenge. 
Thus, such investigations are devoted to the future research. 
\section*{Acknowledgements}
%
KM's work was 
done within the framework of the project from the Polish Science Center 
(NCN) Grant No. 2020/37/B/ST9/00184. 
We gratefully acknowledge Poland’s high-performance computing infrastructure PLGrid (HPC Centers: ACK Cyfronet AGH) for providing computer facilities and support within computational grant no. PLG/2022/015868. 
A part of numerical simulations was run on the LUNAR cluster at the Institute of Mathematics at M. Curie-Skłodowska University in Lublin, Poland. AKS acknowledges RAC-ISRO grant for the support of his scientific research.
We visualize the simulation data using 
the ViSIt software package \citep{Childs_et_al_2012} and 
Python scripts.

\section*{Data availability}
All data underlying the results are available in the article and no additional source data are required.

\bibliographystyle{mnras} 
\bibliography{draft}

\begin{thebibliography}{}
\makeatletter
\relax
\def\mn@urlcharsother{\let\do\@makeother \do\$\do\&\do\#\do\^\do\_\do\%\do\~}
\def\mn@doi{\begingroup\mn@urlcharsother \@ifnextchar [ {\mn@doi@} {\mn@doi@[]}}
\def\mn@doi@[#1]#2{\def\@tempa{#1}\ifx\@tempa\@empty \href {http://dx.doi.org/#2} {doi:#2}\else \href {http://dx.doi.org/#2} {#1}\fi \endgroup}
\def\mn@eprint#1#2{\mn@eprint@#1:#2::\@nil}
\def\mn@eprint@arXiv#1{\href {http://arxiv.org/abs/#1} {{\tt arXiv:#1}}}
\def\mn@eprint@dblp#1{\href {http://dblp.uni-trier.de/rec/bibtex/#1.xml} {dblp:#1}}
\def\mn@eprint@#1:#2:#3:#4\@nil{\def\@tempa {#1}\def\@tempb {#2}\def\@tempc {#3}\ifx \@tempc \@empty \let \@tempc \@tempb \let \@tempb \@tempa \fi \ifx \@tempb \@empty \def\@tempb {arXiv}\fi \@ifundefined {mn@eprint@\@tempb}{\@tempb:\@tempc}{\expandafter \expandafter \csname mn@eprint@\@tempb\endcsname \expandafter{\@tempc}}}

\bibitem[\protect\citeauthoryear{{Avrett} \& {Loeser}}{{Avrett} \& {Loeser}}{2008}]{2008ApJS..175..229A}
{Avrett} E.~H.,  {Loeser} R.,  2008, \mn@doi [\apjs] {10.1086/523671}, \href {https://ui.adsabs.harvard.edu/abs/2008ApJS..175..229A} {175, 229}

\bibitem[\protect\citeauthoryear{{Ballester} et~al.,}{{Ballester} et~al.}{2018}]{2018SSRv..214...58B}
{Ballester} J.~L.,  et~al., 2018, \mn@doi [\ssr] {10.1007/s11214-018-0485-6}, \href {https://ui.adsabs.harvard.edu/abs/2018SSRv..214...58B} {214, 58}

\bibitem[\protect\citeauthoryear{Childs et~al.,}{Childs et~al.}{2012}]{Childs_et_al_2012}
Childs H.,  et~al., 2012, {VisIt: An End-User Tool For Visualizing and Analyzing Very Large Data}.
Chapman and Hall/CRC

\bibitem[\protect\citeauthoryear{{Erd{\'e}lyi} \& {James}}{{Erd{\'e}lyi} \& {James}}{2004}]{2004A&A...427.1055E}
{Erd{\'e}lyi} R.,  {James} S.~P.,  2004, \mn@doi [\aap] {10.1051/0004-6361:20040345}, \href {https://ui.adsabs.harvard.edu/abs/2004A&A...427.1055E} {427, 1055}

\bibitem[\protect\citeauthoryear{{Fleck}, {Carlsson}, {Khomenko}, {Rempel}, {Steiner}  \& {Vigeesh}}{{Fleck} et~al.}{2021}]{2021RSPTA.37900170F}
{Fleck} B.,  {Carlsson} M.,  {Khomenko} E.,  {Rempel} M.,  {Steiner} O.,   {Vigeesh} G.,  2021, \mn@doi [Philosophical Transactions of the Royal Society of London Series A] {10.1098/rsta.2020.0170}, \href {https://ui.adsabs.harvard.edu/abs/2021RSPTA.37900170F} {379, 20200170}

\bibitem[\protect\citeauthoryear{{Forteza}, {Oliver}, {Ballester}  \& {Khodachenko}}{{Forteza} et~al.}{2007}]{2007A&A...461..731F}
{Forteza} P.,  {Oliver} R.,  {Ballester} J.~L.,   {Khodachenko} M.~L.,  2007, \mn@doi [\aap] {10.1051/0004-6361:20065900}, \href {https://ui.adsabs.harvard.edu/abs/2007A&A...461..731F} {461, 731}

\bibitem[\protect\citeauthoryear{{Hansteen}, {De Pontieu}, {Rouppe van der Voort}, {van Noort}  \& {Carlsson}}{{Hansteen} et~al.}{2006}]{2006ApJ...647L..73H}
{Hansteen} V.~H.,  {De Pontieu} B.,  {Rouppe van der Voort} L.,  {van Noort} M.,   {Carlsson} M.,  2006, \mn@doi [\apjl] {10.1086/507452}, \href {https://ui.adsabs.harvard.edu/abs/2006ApJ...647L..73H} {647, L73}

\bibitem[\protect\citeauthoryear{{Heggland}, {Hansteen}, {De Pontieu}  \& {Carlsson}}{{Heggland} et~al.}{2011}]{2011ApJ...743..142H}
{Heggland} L.,  {Hansteen} V.~H.,  {De Pontieu} B.,   {Carlsson} M.,  2011, \mn@doi [\apj] {10.1088/0004-637X/743/2/142}, \href {https://ui.adsabs.harvard.edu/abs/2011ApJ...743..142H} {743, 142}

\bibitem[\protect\citeauthoryear{{Kayshap}, {Murawski}, {Srivastava}, {Musielak}  \& {Dwivedi}}{{Kayshap} et~al.}{2018}]{2018MNRAS.479.5512K}
{Kayshap} P.,  {Murawski} K.,  {Srivastava} A.~K.,  {Musielak} Z.~E.,   {Dwivedi} B.~N.,  2018, \mn@doi [\mnras] {10.1093/mnras/sty1861}, \href {https://ui.adsabs.harvard.edu/abs/2018MNRAS.479.5512K} {479, 5512}

\bibitem[\protect\citeauthoryear{{Khomenko}, {Collados}, {D{\'\i}az}  \& {Vitas}}{{Khomenko} et~al.}{2014}]{2014PhPl...21i2901K}
{Khomenko} E.,  {Collados} M.,  {D{\'\i}az} A.,   {Vitas} N.,  2014, \mn@doi [Physics of Plasmas] {10.1063/1.4894106}, \href {https://ui.adsabs.harvard.edu/abs/2014PhPl...21i2901K} {21, 092901}

\bibitem[\protect\citeauthoryear{{Ku{\'z}ma}, {Kadowaki}, {Murawski}, {Musielak}, {Poedts}, {Yuan}  \& {Feng}}{{Ku{\'z}ma} et~al.}{2024}]{kuzma2024RSPTA.38230218K}
{Ku{\'z}ma} B.,  {Kadowaki} L. H.~S.,  {Murawski} K.,  {Musielak} Z.~E.,  {Poedts} S.,  {Yuan} D.,   {Feng} X.,  2024, \mn@doi [Philosophical Transactions of the Royal Society of London Series A] {10.1098/rsta.2023.0218}, \href {https://ui.adsabs.harvard.edu/abs/2024RSPTA.38230218K} {382, 20230218}

\bibitem[\protect\citeauthoryear{{Li}, {Xu}, {eng}, {Xie}, {Shi}  \& {Deng}}{{Li} et~al.}{2024}]{2024arXiv240102617L}
{Li} K.~J.,  {Xu} J.~C.,  {eng} W.~F.,  {Xie} J.~L.,  {Shi} X.~J.,   {Deng} L.~H.,  2024, \mn@doi [arXiv e-prints] {10.48550/arXiv.2401.02617}, \href {https://ui.adsabs.harvard.edu/abs/2024arXiv240102617L} {p. arXiv:2401.02617}

\bibitem[\protect\citeauthoryear{{Mart{\'\i}nez-Sykora}, {De Pontieu}, {Hansteen}, {Rouppe van der Voort}, {Carlsson}  \& {Pereira}}{{Mart{\'\i}nez-Sykora} et~al.}{2017}]{2017Sci...356.1269M}
{Mart{\'\i}nez-Sykora} J.,  {De Pontieu} B.,  {Hansteen} V.~H.,  {Rouppe van der Voort} L.,  {Carlsson} M.,   {Pereira} T.~M.~D.,  2017, \mn@doi [Science] {10.1126/science.aah5412}, \href {https://ui.adsabs.harvard.edu/abs/2017Sci...356.1269M} {356, 1269}

\bibitem[\protect\citeauthoryear{{Mart{\'\i}nez-Sykora}, {Szydlarski}, {Hansteen}  \& {De Pontieu}}{{Mart{\'\i}nez-Sykora} et~al.}{2020}]{2020ApJ...900..101M}
{Mart{\'\i}nez-Sykora} J.,  {Szydlarski} M.,  {Hansteen} V.~H.,   {De Pontieu} B.,  2020, \mn@doi [\apj] {10.3847/1538-4357/ababa3}, \href {https://ui.adsabs.harvard.edu/abs/2020ApJ...900..101M} {900, 101}

\bibitem[\protect\citeauthoryear{{Murawski}, {Musielak}, {Poedts}, {Srivastava}  \& {Kadowaki}}{{Murawski} et~al.}{2022}]{2022Ap&SS.367..111M}
{Murawski} K.,  {Musielak} Z.~E.,  {Poedts} S.,  {Srivastava} A.~K.,   {Kadowaki} L.,  2022, \mn@doi [\apss] {10.1007/s10509-022-04152-4}, \href {https://ui.adsabs.harvard.edu/abs/2022Ap&SS.367..111M} {367, 111}

\bibitem[\protect\citeauthoryear{{Niedziela}, {Murawski}, {Kadowaki}, {Zaqarashvili}  \& {Poedts}}{{Niedziela} et~al.}{2022}]{2022A&A...668A..32N}
{Niedziela} R.,  {Murawski} K.,  {Kadowaki} L.,  {Zaqarashvili} T.,   {Poedts} S.,  2022, \mn@doi [\aap] {10.1051/0004-6361/202244844}, \href {https://ui.adsabs.harvard.edu/abs/2022A&A...668A..32N} {668, A32}

\bibitem[\protect\citeauthoryear{{Parnell}}{{Parnell}}{2002}]{2002MNRAS.335..389P}
{Parnell} C.~E.,  2002, \mn@doi [\mnras] {10.1046/j.1365-8711.2002.05618.x}, \href {https://ui.adsabs.harvard.edu/abs/2002MNRAS.335..389P} {335, 389}

\bibitem[\protect\citeauthoryear{{Pelekhata}, {Murawski}  \& {Poedts}}{{Pelekhata} et~al.}{2023}]{2023A&A...669A..47P}
{Pelekhata} M.,  {Murawski} K.,   {Poedts} S.,  2023, \mn@doi [\aap] {10.1051/0004-6361/202244671}, \href {https://ui.adsabs.harvard.edu/abs/2023A&A...669A..47P} {669, A47}

\bibitem[\protect\citeauthoryear{{Proctor}}{{Proctor}}{2004}]{2004A&G....45d..14P}
{Proctor} M.~R.~E.,  2004, \mn@doi [Astronomy and Geophysics] {10.1046/j.1468-4004.2003.45414.x}, \href {https://ui.adsabs.harvard.edu/abs/2004A&G....45d..14P} {45, 4.14}

\bibitem[\protect\citeauthoryear{{Routh}, {Musielak}, {Sundar}, {Joshi}  \& {Charan}}{{Routh} et~al.}{2020}]{2020Ap&SS.365..139R}
{Routh} S.,  {Musielak} Z.~E.,  {Sundar} M.~N.,  {Joshi} S.~S.,   {Charan} S.,  2020, \mn@doi [\apss] {10.1007/s10509-020-03852-z}, \href {https://ui.adsabs.harvard.edu/abs/2020Ap&SS.365..139R} {365, 139}

\bibitem[\protect\citeauthoryear{{Srivastava} et~al.,}{{Srivastava} et~al.}{2021}]{2021JGRA..12629097S}
{Srivastava} A.~K.,  et~al., 2021, \mn@doi [Journal of Geophysical Research (Space Physics)] {10.1029/2020JA029097}, \href {https://ui.adsabs.harvard.edu/abs/2021JGRA..12629097S} {126, e029097}

\bibitem[\protect\citeauthoryear{{Tian}, {Tu}, {Marsch}, {He}  \& {Kamio}}{{Tian} et~al.}{2010}]{2010ApJ...709L..88T}
{Tian} H.,  {Tu} C.,  {Marsch} E.,  {He} J.,   {Kamio} S.,  2010, \mn@doi [\apjl] {10.1088/2041-8205/709/1/L88}, \href {https://ui.adsabs.harvard.edu/abs/2010ApJ...709L..88T} {709, L88}

\bibitem[\protect\citeauthoryear{{Tu}, {Zhou}, {Marsch}, {Xia}, {Zhao}, {Wang}  \& {Wilhelm}}{{Tu} et~al.}{2005}]{2005Sci...308..519T}
{Tu} C.-Y.,  {Zhou} C.,  {Marsch} E.,  {Xia} L.-D.,  {Zhao} L.,  {Wang} J.-X.,   {Wilhelm} K.,  2005, \mn@doi [Science] {10.1126/science.1109447}, \href {https://ui.adsabs.harvard.edu/abs/2005Sci...308..519T} {308, 519}

\bibitem[\protect\citeauthoryear{{Vigeesh}, {Jackiewicz}  \& {Steiner}}{{Vigeesh} et~al.}{2017}]{2017ApJ...835..148V}
{Vigeesh} G.,  {Jackiewicz} J.,   {Steiner} O.,  2017, \mn@doi [\apj] {10.3847/1538-4357/835/2/148}, \href {https://ui.adsabs.harvard.edu/abs/2017ApJ...835..148V} {835, 148}

\bibitem[\protect\citeauthoryear{{Wi{\'s}niewska}, {Musielak}, {Staiger}  \& {Roth}}{{Wi{\'s}niewska} et~al.}{2016}]{2016ApJ...819L..23W}
{Wi{\'s}niewska} A.,  {Musielak} Z.~E.,  {Staiger} J.,   {Roth} M.,  2016, \mn@doi [\apjl] {10.3847/2041-8205/819/2/L23}, \href {https://ui.adsabs.harvard.edu/abs/2016ApJ...819L..23W} {819, L23}

\bibitem[\protect\citeauthoryear{{W{\'o}jcik}, {Murawski}  \& {Musielak}}{{W{\'o}jcik} et~al.}{2018}]{2018MNRAS.481..262W}
{W{\'o}jcik} D.,  {Murawski} K.,   {Musielak} Z.~E.,  2018, \mn@doi [\mnras] {10.1093/mnras/sty2306}, \href {https://ui.adsabs.harvard.edu/abs/2018MNRAS.481..262W} {481, 262}

\bibitem[\protect\citeauthoryear{{W{\'o}jcik}, {Ku{\'z}ma}, {Murawski}  \& {Musielak}}{{W{\'o}jcik} et~al.}{2020}]{2020A&A...635A..28W}
{W{\'o}jcik} D.,  {Ku{\'z}ma} B.,  {Murawski} K.,   {Musielak} Z.~E.,  2020, \mn@doi [\aap] {10.1051/0004-6361/201936938}, \href {https://ui.adsabs.harvard.edu/abs/2020A&A...635A..28W} {635, A28}

\bibitem[\protect\citeauthoryear{{Zaqarashvili}, {Khodachenko}  \& {Rucker}}{{Zaqarashvili} et~al.}{2011}]{2011A&A...529A..82Z}
{Zaqarashvili} T.~V.,  {Khodachenko} M.~L.,   {Rucker} H.~O.,  2011, \mn@doi [\aap] {10.1051/0004-6361/201016326}, \href {https://ui.adsabs.harvard.edu/abs/2011A&A...529A..82Z} {529, A82}

\bibitem[\protect\citeauthoryear{{del Pino Alem{\'a}n}, {Trujillo Bueno}, {{\v{S}}t{\v{e}}p{\'a}n}  \& {Shchukina}}{{del Pino Alem{\'a}n} et~al.}{2018}]{2018ApJ...863..164D}
{del Pino Alem{\'a}n} T.,  {Trujillo Bueno} J.,  {{\v{S}}t{\v{e}}p{\'a}n} J.,   {Shchukina} N.,  2018, \mn@doi [\apj] {10.3847/1538-4357/aaceab}, \href {https://ui.adsabs.harvard.edu/abs/2018ApJ...863..164D} {863, 164}

\makeatother
\end{thebibliography}

\end{document}